\documentclass[useAMS,usenatbib]{mn2e}

\usepackage{lscape,graphicx,pifont}
\usepackage{times}
\usepackage{color}
\usepackage{longtable}
\usepackage{natbib}

\DeclareGraphicsExtensions{.png,.pdf,.eps,.jpg,.tiff,.ps}

\newcommand{\Msun}{\mbox{\,M$_\odot$}}
\newcommand{\Lsun}{\mbox{\,L$_\odot$}}
\newcommand{\vunit}{\mbox{\,km\,s$^{-1}$}}
\newcommand{\mic}{\mbox{$\,\mu$m}}
\newcommand{\pion}[2]{{#1}\,{\sc {#2}}}
\newcommand{\fion}[2]{[{#1}\,{\sc {#2}}]}

\newcommand{\ltsimeq}{\raisebox{-0.6ex}{$\,\stackrel
        {\raisebox{-.2ex}{$\textstyle <$}}{\sim}\,$}}

\newcommand{\sirtf}{\mbox{\it Spitzer}}
\newcommand{\spitzer}{\mbox{\it Spitzer Space Telescope}}

\newcommand{\sof}{\mbox{\it SOFIA}}

\newcommand{\tcrb}{\mbox{T~CrB}}
\newcommand{\rs}{\mbox{RS~Oph}}

\title[The environment of T~CrB]{Gas phase SiO in the circumstellar environment 
of the recurrent nova T~Coronae Borealis}

\author[A. Evans et al.]{A. Evans$^{1}$\thanks{E-mail: a.evans@keele.ac.uk},
Ya. V. Pavlenko$^{2,3}$,
D. P. K. Banerjee$^4$,
U. Munari$^5$,
R. D. Gehrz$^6$,\newauthor
C. E. Woodward$^6$, 
S. Starrfield$^7$,
L. A. Helton$^8$,
M. Shahbandeh$^9$, S. Davis$^9$, \newauthor
S. Dallaporta$^{10}$, G. Cherini$^{10}$\\ \\
{$^1$}Astrophysics Group, Lennard Jones Laboratory, Keele University, Keele, Staffordshire, 
ST5 5BG, UK\\
{$^2$}Main Astronomical Observatory, Academy of Sciences of the Ukraine, Golosiiv Woods, Kyiv-127, 03680 Ukraine\\
{$^3$}Centre for Astrophysics Research, University of Hertfordshire, College Lane, Hatfield, Hertfordshire AL10 9AB, UK\\
{$^4$}{Physical Research Laboratory, Navrangpura,  Ahmedabad, Gujarat 380009, India}\\
{$^5$}INAF Astronomical Observatory of Padova, I-36012 Asiago (VI), Italy\\
{$^6$}Minnesota Institute for Astrophysics, School of Physics \& Astronomy,
116 Church Street SE, University of Minnesota, Minneapolis, MN 55455, USA\\
$^{7}$School of Earth and Space Exploration, Arizona State University, Box 871404, Tempe, 
AZ 85287-1404, USA\\ 
$^{8}$USRA-SOFIA Science Center, NASA Ames Research Center, Moffett Field, CA 94035, USA\\ 
${^9}$Department of Physics, Florida State University, 77 Chieftan Way, Tallahassee, FL 32306, USA \\ 
$^{10}$ANS Collaboration, Astronomical Observatory, 36012 Asiago (VI), Italy
}

\begin{document}

\date{Version of \today}

\pagerange{\pageref{firstpage}--\pageref{lastpage}} \pubyear{2019}

\maketitle

\label{firstpage}

\begin{abstract}
 We report the discovery of 
 the diatomic molecule SiO in the gas phase in the environment
 of the recurrent nova T~Coronae Borealis. 
While some of the SiO is photospheric,
a substantial portion must arise in the wind from the red giant component of \tcrb.
A simple fit to the SiO feature, assuming local thermodynamic equilibrium,
suggests a SiO column density of 
{$2.8\times10^{17}$~cm$^{-2}$} and temperature
$\sim1,000$~K; the SiO column density is 
{similar to that present in the winds of field red giants.}
A search for SiO maser emission is encouraged both 
before and after the next anticipated eruption.
We find that the $^{12}$C/$^{13}$C ratio in the red giant 
{is $<9$, with a best fit value of $\sim5$, a} factor $\sim18$ times 
lower than the solar value of 89.
We find no convincing evidence for the presence of dust in the environment 
of \tcrb, which we attribute to the destructive effects on nucleation sites 
of hard X-ray emission. When the next eruption of \tcrb\ occurs, the 
ejected material will shock the wind, 
producing X-ray and coronal line emission, as is the
case for the recurrent nova \rs. \tcrb\ is also a good candidate for 
very high energy $\gamma$-ray emission, as first observed during the 
2010 outburst of V407~Cyg.
 {We include in the paper a wide variety of infrared spectroscopic and photometric data.}

\end{abstract}



\section{Introduction}
\label{intro}

Recurrent novae (RNe) are a subclass of the cataclysmic variables 
that consist
of a compact component (a white dwarf; WD) and a cool donor star. 
Depending on the nature of the system, material from the donor star 
spills over onto the WD via an accretion disc.
 
Like classical novae, they are the result of a thermonuclear runaway on 
the surface of the WD. All novae are  recurrent but RN eruptions, unlike 
those of classical novae, recur on a human time-scale ($\ltsimeq100$~yrs)
rather than an astronomical time-scale \citep{starrfield}.
Therefore, RNe are defined by the selection effect that they
have been observed to undergo more than one eruption. 

There is strong evidence that the mass of the white dwarf (WD) in RN systems
is close to the Chandrasekhar limit \citep[e.g][]{thgood}. As material from the 
donor star spills over onto the WD, it is possible that the net flow 
of matter onto the WD (taking into account any mass lost in eruptions) is positive, 
the mass of the WD may increase, eventually taking it over the Chandrasekhar 
limit \citep[e.g.][]{starrfield2}. 
 
There are undoubtedly several channels to Type~Ia supernova explosions, 
although none is currently clearly favoured  \citep[see][and references therein]{maoz}.
However, should the WD in a RN system gain mass, and it has Carbon-Oxygen 
composition \citep{maoz}, such RN systems may be one of the progenitors of 
Type~Ia supernovae \citep[see][]{MR,maoz}. Studying RNe therefore provides 
us with the opportunity to understand the progenitors of such supernovae, 
which are a key tool in the determination of the cosmic distance scale and of 
cosmic large-scale structure \citep{perlmutter-1,perlmutter-2,riess}.
 
RNe are a heterogeneous set of objects but they fall naturally into two 
types \citep{anupama}: (a)~those with long orbital
periods (e.g. \rs, \tcrb, orbital periods of months--years), in which the cool 
component is a red giant (RG) and (b)~those with short orbital periods (e.g. 
U~Sco, T~Pyx, orbital periods of hours--days);  the cool component in these 
systems may be a main sequence dwarf or a subgiant.
 
\section{Long period recurrent nova systems}
\label{long}
 
Four RN systems with long orbital periods are (as of 2019) known, 
although only \tcrb\ and \rs\ have well-determined orbital periods 
\citep[227.67~d and 455.72~d respectively; see][]{anupama}; 
the long orbital periods of the other two 
(V745~Sco and V3890~Sgr)
are inferred from the fact that they have RG secondaries. 
Such systems are commonly referred to as ``symbiotic binaries''.

In these long-period systems, the RG usually has a wind and, when the RN
eruption occurs, the ejecta run into and shock the RG wind. 
The subsequent deceleration of the ejecta is manifested in the narrowing 
of emission lines as the eruption proceeds \citep[e.g.][]{evans3,munari3}.
Hydrodynamical modelling of this phenomenon has been carried out 
by \cite*{walder} for the case of the 2006 eruption of \rs, 
and by \cite*{pan} for the 2010 eruption of the symbiotic binary V407~Cyg.
The interaction between 
the ejecta and the wind results in X-ray and non-thermal radio emission, as 
well as coronal line emission in the optical and infrared (IR); 
very high energy $\gamma$-ray emission -- as seen in V407~Cyg 
\citep{cheung} -- may also be expected  \citep{banerjee}.

The outbursts of the long period systems are very 
homogeneous, to the extent that the visual light curves of successive eruptions 
of a given RN are practically indistinguishable \citep{anupama}. They behave 
like ``fast'' novae of the He/N type: the light curves decline at 
$\sim0.3$~mag~day$^{-1}$, and material is ejected at high velocity, several 
thousand \vunit. The 2006 eruption of \rs\ was particularly well observed, 
from X-rays to radio \citep*{das,evans-asp,banerjee2,drake,ness}. 


\section{The recurrent nova \tcrb}
\label{tcrb}

\tcrb\ has undergone eruptions in 1866 and 1946, and
there may have been an eruption in 1842 \citep[][but see also \cite{schaefer3}]{schaefer}.
The properties of \tcrb\ are discussed in detail by \cite{anupama1}; 
an updated account is given by \cite*{munari}. 

\cite{anupama} gives its distance and interstellar reddening as 1.3~kpc 
and $E(B-V)=0.15$ respectively. \cite{schaefer2} gives a distance 
of $800\pm140$~pc, by an average of several independent
methods. While there is a parallax for \tcrb\ from the 
{\it Gaia} survey (leading to a distance of $825\pm33$~pc), this 
may be unreliable because of orbital motion \citep{schaefer1}.
The reddening is likely somewhat smaller than \citeauthor{anupama}'s 
value \citep*[see][]{schlegel,schlafly,green}. According to \citeauthor{green},
the reddening in the direction of \tcrb\ levels out to a value $E(B-V)=0.07\pm0.02$ 
beyond $\sim350$~pc; \citeauthor{schlafly} give $E(B-V)=0.058$, while 
\citeauthor{schlegel} give $E(B-V)=0.062$. We assume $E(B-V)=0.06$ here. 
The reddening is so low that it is a 
very minor source of uncertainty in our analysis.

\begin{figure}
\includegraphics[width=8cm,keepaspectratio]{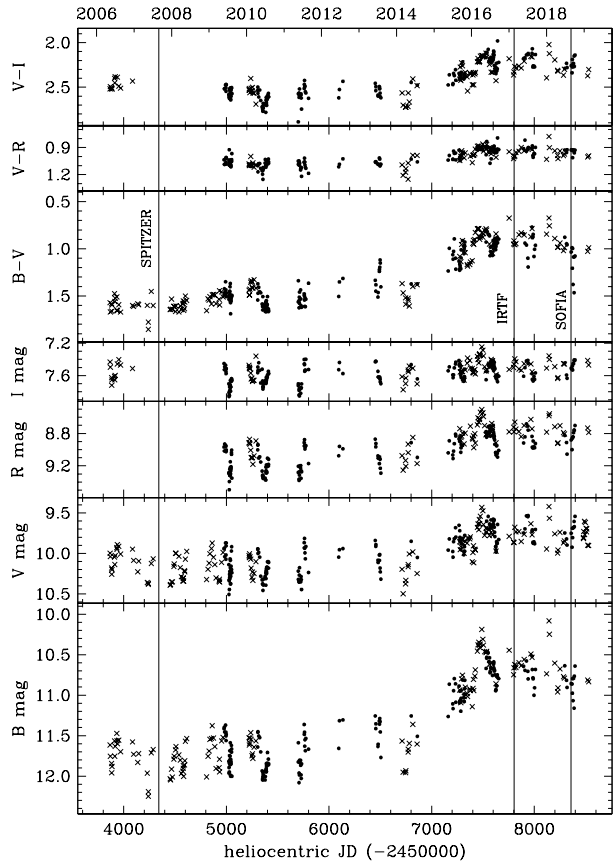}
\caption{$BV\!RI$ light curves of \tcrb\ covering the \sirtf, IRTF and \sof\ 
observations; the JD times of the \sirtf, IRTF and \sof\ observations are indicated.
Data from the ANS Collaboration; the different symbols represent data from 
two separate telescopes, working independently to provide an independent check.
\label{tcrb_lc}}
\end{figure}

\begin{figure*}
\includegraphics[width=12cm,keepaspectratio]{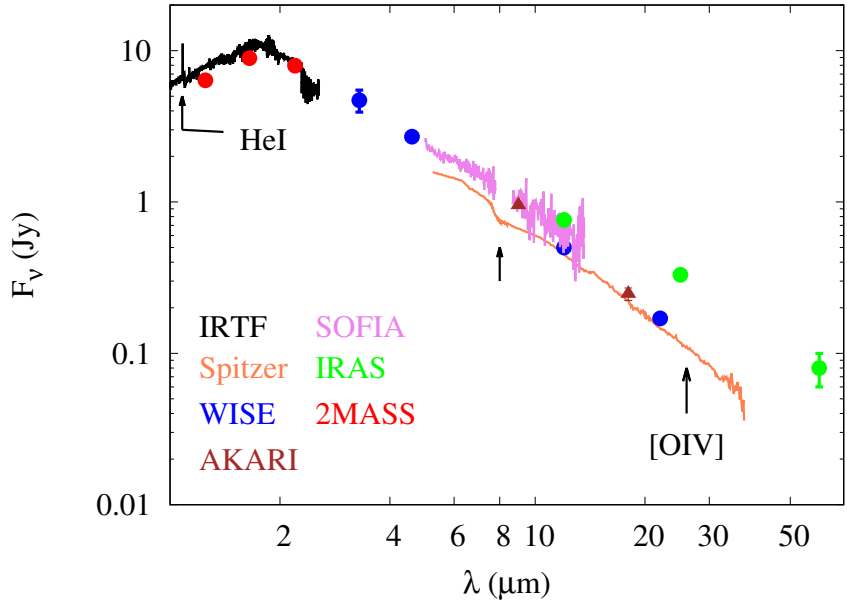}
\caption{IRTF, \sirtf\ IRS and \sof\ FORCAST spectra of \tcrb, with 2MASS, {\it WISE},
{\it AKARI} and 
{\it IRAS} photometry superimposed; errors in 2MASS, {\it WISE}, {\it AKARI} and 
{\it IRAS} photometry are smaller than points if not visible.
The \pion{He}{i} $\lambda=1.083$\mic\ line is labelled, as is the expected location of the
\fion{O}{iv} $\lambda=25.89$\mic\ line.
The arrow at 8\mic\ shows the SiO absorption feature discussed in Section~\ref{SiO_disc}.
\label{tcrb_sp}}
\end{figure*}

The post-1946 photometric evolution is described by \cite{munari}, who find 
evidence for a super-active photometric phase starting in 2015. This 
is well illustrated in Fig.~\ref{tcrb_lc}, which draws on data from the
Asiago Novae and Symbiotic stars (ANS) collaboration 
\citep{munari-ans,munari-ans2}. 
The apparent scatter in the light curves is real, and is a consequence of the strong 
orbital modulation induced by irradiation of the M3III RG secondary by the WD, 
and ellipsoidal variations due to the RG filling its Roche lobe \citep{YM}. 
The onset of the active state in 2015, which is evident in Fig.~\ref{tcrb_lc},
was accompanied by significant changes in the X-ray emission \citep{luna2}.
There is also evidence for flickering activity \citep{ilkiewicz}.

Both the ultra-violet (UV) continuum and emission lines are highly variable 
\citep{anupama1}. \cite{selvelli} estimated the UV luminosity to be $\sim40$\Lsun, 
implying a mass accretion rate onto the WD of $\sim∼2.5\times 10^{-8}$\Msun~y$^{-1}$.

\tcrb\ was detected by {\it Swift} in 2005 Jun-Oct \citep{kennea}; they find that its 
spectral energy distribution (SED)  in the 0.3--150~keV range
is well fitted by a single-temperature ($\sim28$~keV) bremsstrahlung model. 
The absorbing column density is $\sim21.5\times10^{22}$~cm$^{-2}$, some 
400 times larger than the column density expected from absorption in the general 
interstellar medium. \citeauthor{kennea} conclude that 
the absorption must be intrinsic to \tcrb. They also estimate that the mass 
of the WD in the \tcrb\ system is 1.35\Msun, concluding that the 
WD in \tcrb\ is close to the Chandrasekhar limit.
This finding is consistent with the $M_{\rm WD}=1.3-2.5$\Msun\ derived
by \cite{shahbaz1}, based on modelling the ellipsoidal 
variations in the IR light curve and assuming an orbital 
inclination of order $38^\circ-46^\circ$. \cite*{luna1} observed 
\tcrb\ with the {\it Suzaku} 
X-ray observatory in 2006 September. They too found that there is a substantial 
($\sim35\times10^{22}$~cm$^{-2}$) circumstellar hydrogen column density.

While symbiotic binaries are known to be soft X-ray sources,
\tcrb\ is one of a small number of this class known to be
a hard X-ray source \citep[][and references therein]{eze}; it is
likely that the hard X-rays arise at the boundary layer between the WD
and the accretion disc \citep[e.g.][]{kennea}.

Recently, \cite{shara18} summarised the WD masses and 
accretion rates for several recurrent and classical novae. For 
\tcrb\ they give $\sim1.35\Msun$ for the WD mass and  
$2.1\times10^{-8}$\Msun~y$^{-1}$ for the accretion rate, suggesting 
that both parameters are well determined for \tcrb.

\cite{luna3} recently described {\it Swift} and {\it XMM}-Newton 
observations of \tcrb, combined with optical data over a 12 year period. 
These data show a marked rise in the optical/UV  flux accompanied by a 
decline in the X-ray flux. \citeauthor{luna3} ascribe this to a disc 
instability event, and that the optical brightening is similar to that 
seen some 8~years before the 1946 eruption. This is consistent with 
the analysis by \cite{munari}, who suggest that \tcrb\ is due to erupt 
around 2026. \citeauthor{luna3} estimate an X-ray absorbing column density
of $\sim6.8\times10^{23}$~cm$^{-2}$, significantly higher than is the 
case for the earlier X-ray observations \citep{kennea,luna1}.

It is important that the \tcrb\ system is well observed prior to 
the next eruption so that the evolution of the eruption can be interpreted
in an informed manner. In this paper we present IR observations of \tcrb,
and report the discovery of gas-phase molecular SiO in absorption, and 
{an estimation} of the $^{12}$C/$^{13}$C ratio in the RG component.

\section{Observations}

\subsection{Infrared photometric observations}

There are a number of IR photometric observations of \tcrb\ in the public
domain.
\begin{enumerate}
\item It is listed as an {\it IRAS} source by \cite{harrison}, who 
found fluxes (in Jy) of $0.76\pm0.03$, $0.33\pm0.02$, $0.08\pm0.02$ 
in {\it IRAS} Bands 1 (12\mic), 2 (25\mic) and 3 (60\mic) respectively; 
it was not detected in Band 4 (100\mic) to a limit of 0.31~Jy.
\item It is detected in the Wide-field Infrared Survey Explorer 
\citep[{\it WISE};][]{wise} survey, with fluxes (in Jy)
of 4.70, 2.70, 0.50 and 0.17 in the {\it WISE} bands 1 (3.3\mic),
2 (4.6\mic), 3 (12\mic) and 4 (22\mic) respectively \citep{evans-wise}. 
The {\it WISE} photometry is fitted by a 4,300~K blackbody \citep{evans-wise},
somewhat hotter than the 3,500~K expected for a M3III star \citep{cox}.
\item \tcrb\ also appears in the 2MASS \citep{2mass} survey, with fluxes 
(in Jy) of 6.37 ($J$), 8.91 ($H$), 7.95 ($K_s$).
\item \tcrb\ is an {\it AKARI} \citep{akari} source, with flux densities
$0.95\pm0.04$~Jy at 9\mic\ and $0.25\pm0.02$~Jy at 18\mic.
\end{enumerate}
The 2MASS, {\it WISE}, {\it AKARI} and {\it IRAS} data are plotted in Fig.~\ref{tcrb_sp}.
It is noticeable that the {\it IRAS} fluxes are significantly higher than 
those from {\it WISE}; there is therefore a possibilty 
that \tcrb\ may be variable in the mid-IR on a $\sim20$-year timescale.

\subsection{\spitzer}

\tcrb\ was observed on 2007 August 30 with the \spitzer\ \citep{spitzer,spitzer-g} 
InfraRed Spectrograph \citep[IRS;][]{houck} as part of the \sirtf\ programme 
PID~40060 (P.I. A. Evans).
The \sirtf\ IRS spectrum, shown in Fig.~\ref{tcrb_sp}, reveals the Rayleigh-Jeans 
tail of the RG photosphere; the IRS data are consitent with the {\it WISE} photometry, 
indicating that there has been minimal change in the mid-IR between the \sirtf\ 
(2007) and  {\it WISE} (2011) observations. We note here that both the \sirtf\ and
{\it WISE} data were obtained before the onset of the activity in 2015.

\subsection{NASA Infrared Telescope Facility}

\tcrb\ was observed with the SpeX spectrograph \citep{rayner} on the 3~m NASA 
Infra-Red Telescope Facility
(IRTF) on 2017 February 17.63 UT at spectral 
resolution of 1,200 over the 0.77--2.50\mic\ region.
Details of the observations and data reduction are given in \cite{munari2};
the IRTF data are included in Fig.~\ref{tcrb_sp}. 

\subsection{\sof}

\tcrb\ was observed with the Faint Object IR CAmera for the SOFIA 
Telescope \citep[FORCAST;][]{forcast1,forcast2} instrument on the 
NASA {\it Stratospheric Observatory for Infrared Astronomy} \citep[\sof;][]{sofia}.

The observations were carried out on two separate flights,
on 2018 August 24.24 UT (flight F498) and 2018 August 30.22 UT (flight F494),
as part of our \sof\ Cycle 6 programme to monitor RNe between
eruptions (P.I. A. Evans, proposal ID 06\_0096). 
\tcrb\ was observed with the G063 and G111 grisms, using the
$4.\!\!''7\times191''$ slit to give spectral resolution $R\simeq70$.
The on-target integration times were 1,000~s and 3,000~s for grisms
G063 (flight 498) and G111 (flight 494) respectively.
Both observations used the default 2-point chop/nod ``Nod\_Match\_Chop'' mode.
We assume that, as the time-interval between the two observations
are substantially less than the variations due to orbital modulation,
the two spectral segments can be combined.

The \sof\ data are included in Fig.~\ref{tcrb_sp}.
It seems that the flux density in the 5--13\mic\ range was higher in 
2018 than it was in 2007, when the \sirtf\ data were obtained;
as \tcrb\ is a point source this difference can not be ascribed to
(for example) differences between the instrument parameters
between the \sirtf\ IRS and \sof\ FORCAST, such as slit width. Indeed, while 
the \sirtf\ fluxes are consistent with the {\it WISE} data, the \sof\ fluxes 
are closer to those measured by IRAS in 1983. As we note in Section~\ref{phase} 
below, the orbital phase during both the \sirtf\ and the \sof\
observations were similar: the flux difference between these observations
can not therefore be attributed to either irradiation of the RG or ellipsoidal
variations. It must presumably be due to the increased activity since 2015
(see Fig.~\ref{tcrb_lc}).

\section{Orbital phase}
\label{phase}
Fig.~\ref{tcrb_lc} shows the $BV\!RI$ light curves over the period covered 
by the \sof,  \sirtf\ and IRTF observations. 

As discussed in \cite{munari}, the photospheric temperature of the 
RG changes by up to a couple of spectral sub-types as a result of the 
combined effect of orbital aspect and irradiation by the WD companion.
The orbital phase of the \sirtf\ observation  -- as determined from
the ephemeris of \cite{kenyon} and \cite{fekel} --
was 0.985 (i.e. the irradiated 
side of the RG is hidden from view), while for the IRTF observation it was 
0.680 (i.e. the irradiated face of the RG is almost
in full view). The orbital phase was 0.123
at the time of the \sof\ observation, so again the irradiated face of
the RG was not visible.

The binary phase  must be accounted for in the modelling.

\section{Synthetic spectra}

Synthetic spectra for the RG were computed using the WITA6 program
\citep{pavl97}, assuming local thermodynamic equilibrium (LTE), 
hydrostatic equilibrium and a one-dimensional model atmosphere without 
sources and sinks of
energy. Theoretical synthetic spectra were computed for RG
model atmospheres having effective temperatures in the range
$T_{\rm eff} = 3,000-4,000$~K, 
and gravities $\log{g}$ in the range 0--3 with a gravity step
$\Delta\log{g} =1.0$. A microturbulence velocity of 2\vunit\ was assumed.

In addition to atomic lines, molecular lines 
of H$_2$O, TiO, CrH, VO, CaH, $^{12}$CO and $^{13}$CO, are included 
\citep[see][for details]{PJL,pavl08}. 
Computations were performed for the 1D SAM12 model atmospheres \citep{pavl03}. 
The goodness of fit is determined in the usual way by minimising 
the $S$ parameter \citep{PJL,pavl08}. Direct comparison 
of the SAM12 and MARCS \citep{marcs} model atmospheres shows good 
agreement in their temperature structures ($\ltsimeq50$~K), despite 
the differences in the adopted abundance scales.

We discuss the near-IR and mid-IR separately.

\section{The near-IR}
\subsection{The effects of irradiation}
\label{NIR}
\begin{figure}
\includegraphics[width=8cm]{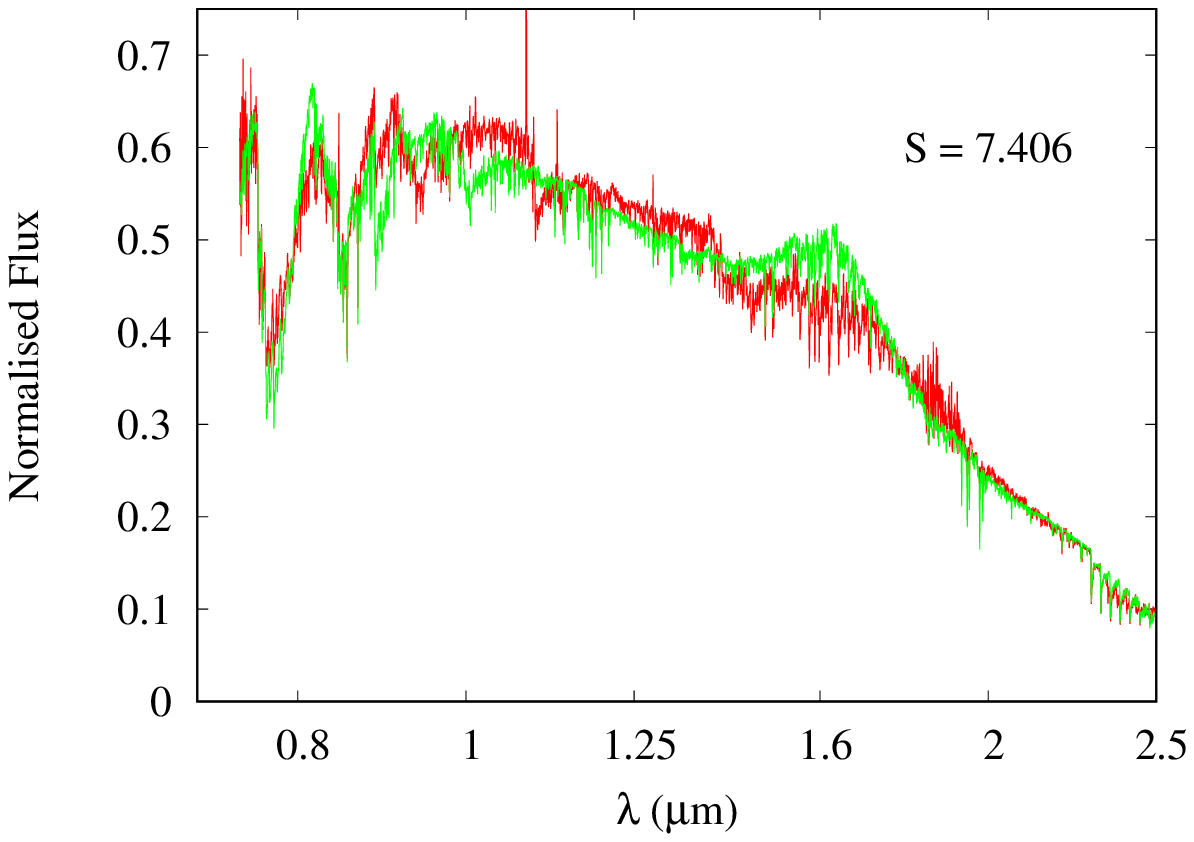}
\includegraphics[width=8cm]{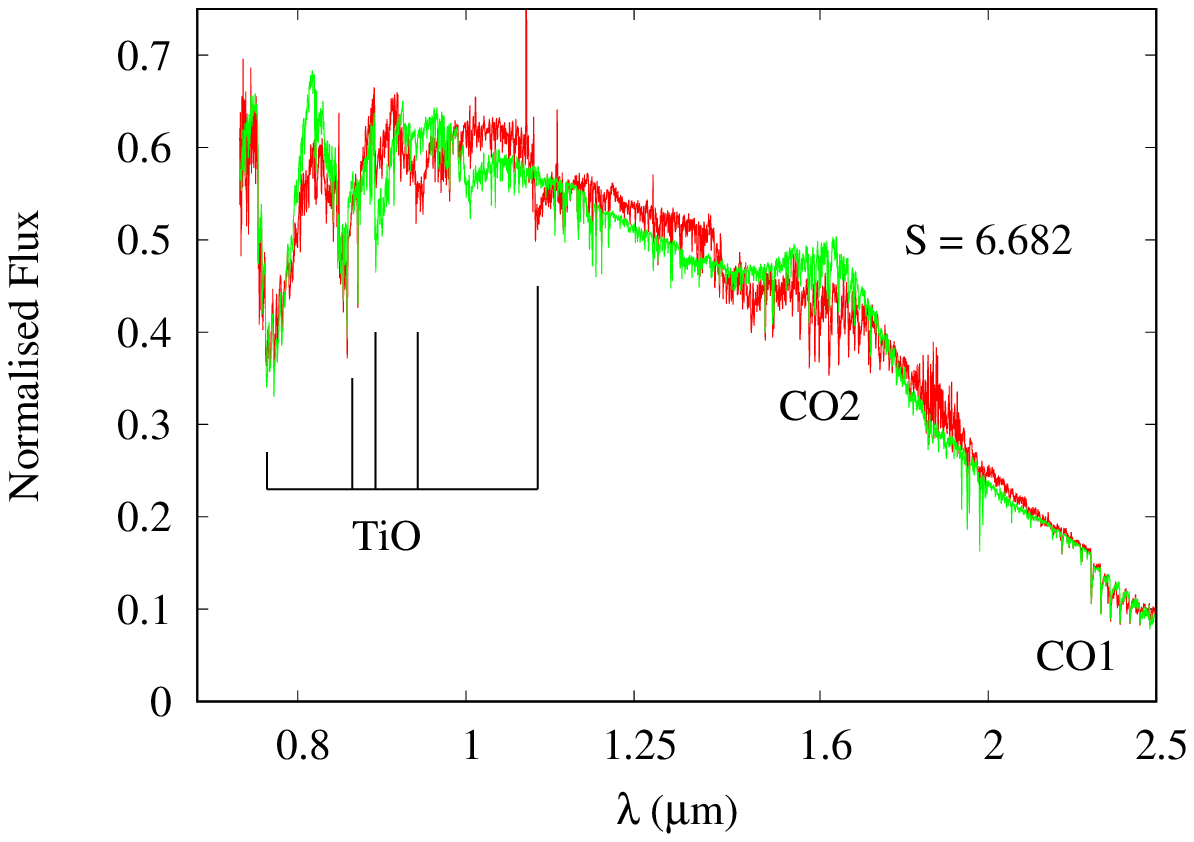}
\caption{Top: fit of synthetic spectrum (green) to the observed near-IR 
SED of T CrB (red); no irradiation.
Bottom: fit of the synthetic spectrum together with irradiation 
by a blackbody as discussed in text.
Wavelength scale is logarithmic to highlight the fit at the shorter
wavelengths. 
In both cases the parameter $S$ represents the goodness of fit.
Bottom frame includes identification of some prominent molecular absorptions;
CO1 and CO2 are the first and second overtone CO bands respectively.
Note that the atmospheric transmission is poor around 1.4\mic\ and 1.8\mic.
\label{irrad}}
\end{figure}

An initial best fit to the near-IR SED
yields a $T_{\rm eff}=3,600$~K, $\log{g}=3$ and solar abundances, as shown
in Fig.~\ref{irrad} (top panel). 

However, we know that
the irradiated face of the RG was in almost full view at the time of the IRTF
observation, and this must be taken into account
\citep[see][for an application of this to the RG in \rs]{pavlenko}.
We find that an improved fit is obtained if some allowance is made for 
the irradiation of the RG by the WD. The model flux is now 
\[ F_{\rm total} = a \, F_{\rm SAM12} + (1-a) \, B_\lambda(T_{\rm BB}) \: ,\]
where $F_{\rm total}$, $F_{\rm SAM12}$ and $B_{\lambda}(T_{\rm BB})$ are the 
total model flux, the flux from the SAM12 synthetic spectrum and the contribution 
of a blackbody at temperature $T_{\rm BB}$, respectively; $(1-a)$ 
effectively represents the fraction of the {visible RG hemisphere
that is irradiated by the WD}. We explored values of $T_{\rm BB}$ of 
5,000~K, 7,350~K, 10,000~K and 12,500~K, and $a$ in the range $0\ge{a}\ge0.9$, 
with increments of 0.1, and $0.9\ge{a}\ge1$, with increments of 0.01.

An improved fit is obtained to the near-IR SED with the same SAM12 
synthetic spectrum, and $T_{\rm BB}=5,000$~K and $a=0.93$ (see bottom 
panel of Fig.~\ref{irrad}).
The calculated $F_{\rm total}$ fits the spectrum well, including the deep 
TiO bands in the red; individual atomic lines are also well reproduced.
{We note in Fig.~\ref{irrad} that there is an apparent excess in the
synthetic spectrum
compared with the data around 1.6\mic. The former is effectively 
for a ``normal'' RG, in which the H$^-$ continuum is prominent, whereas the 
RG in \tcrb\ is in a binary with a WD. The dissociation energy of the H$^-$ ion 
is 0.75~eV, and we suggest that it does not survive, especially on the irradiated 
side of the RG, thus leading to a reduction in the observed flux around 1.6\mic.}

\subsection{The $^{12}$C/$^{13}$C ratio}
\label{1213}
\begin{figure}
         \includegraphics[width=8cm]{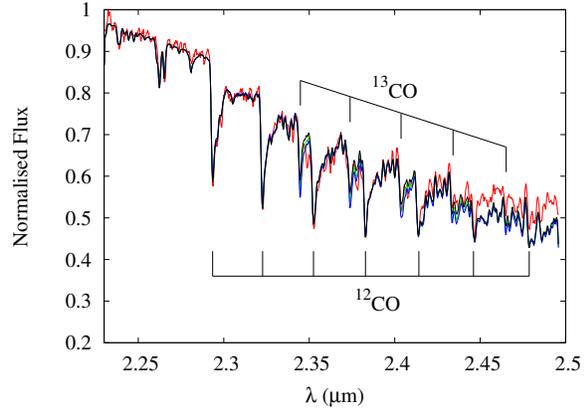}
\caption{{Fit of model fluxes, including irradiation, to the 
observed fluxes (red) across the CO first overtone band with 
$^{12}$C/$^{13}\mbox{C}=5$ (green); other curves are for 
$^{12}$C/$^{13}\mbox{C}=9$ (black) and
$^{12}$C/$^{13}\mbox{C}=3$ (blue).
The $\Delta\upsilon=2$ band-heads are identified.}
\label{12CO}}
\end{figure}

The first overtone CO bands are clearly present in the IRTF spectrum 
(see Fig.~\ref{12CO}); indeed inspection shows that both $^{12}$CO and
$^{13}$CO isotopologues are present.
The fit discussed in Section~\ref{NIR}, including irradiation, applied to 
the 2\mic\ region of the spectrum,
shows that the first overtone CO absorption is best fitted  
with {$^{12}$C/$^{13}\mbox{C}\simeq5$} (see Fig.~\ref{12CO}),
a factor $\sim18$ times lower than the solar value of 89 \citep{wilson,coplen,CN};
{however all the spectral resolution allows us to
conclude is that $^{12}$C/$^{13}\mbox{C}\ltsimeq9$.
Even with this limit the $^{12}$C/$^{13}$\mbox{C} ratio is lower} 
than that found in the RG of \rs\ 
\citep[$^{12}$C/$^{13}\mbox{C}=16\pm3$;][]{pavl10}. Indeed this ratio
in \tcrb\ is towards the extreme low end of the $^{12}$C/$^{13}\mbox{C}$ ratio 
recently found observationally in RGs in the open cluster NGC\,6791 
\citep{szigeti}, and is much lower than that predicted ($\sim20$)
by the post-dredge up evolution scenario \citep{charbonnel}.

\section{The mid-IR}
\subsection{SiO in the RG}
\label{SiO_disc}

\begin{figure}
\includegraphics[width=7.8cm,keepaspectratio]{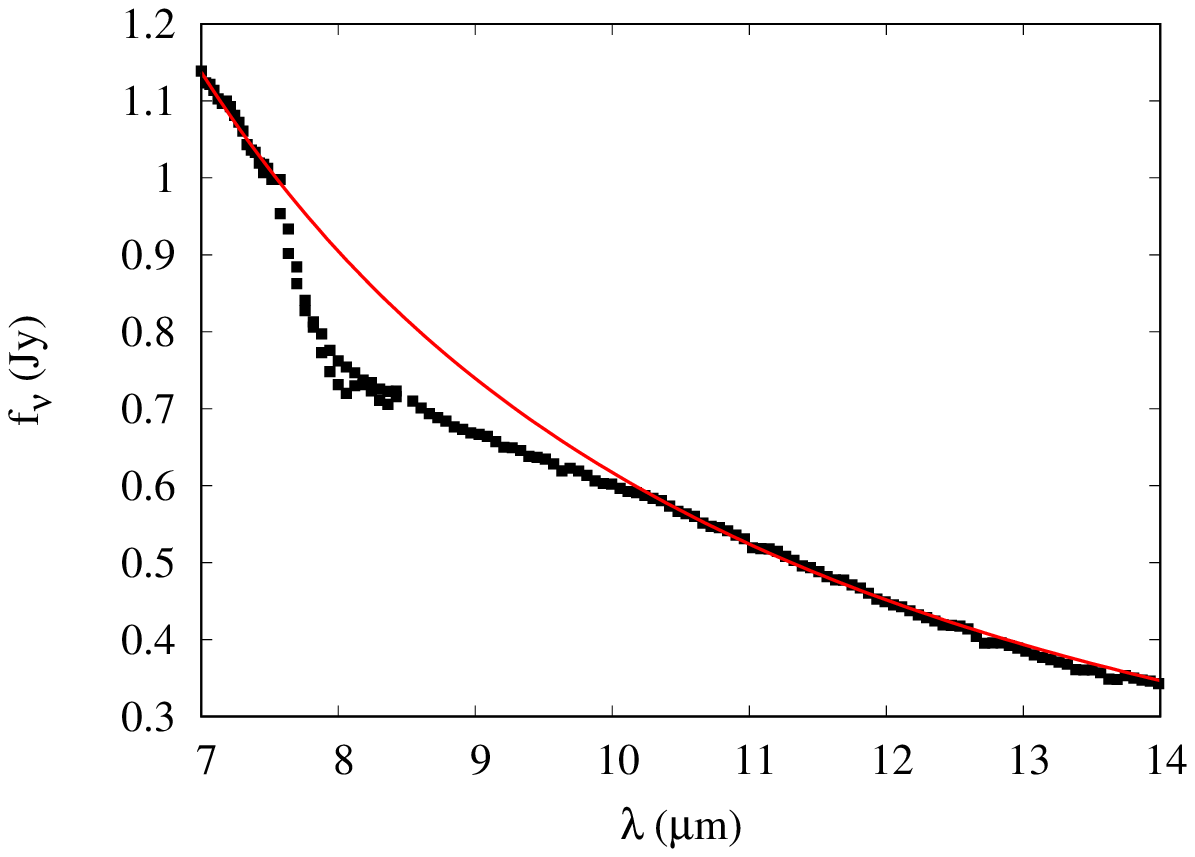}
\includegraphics[width=7.8cm,keepaspectratio]{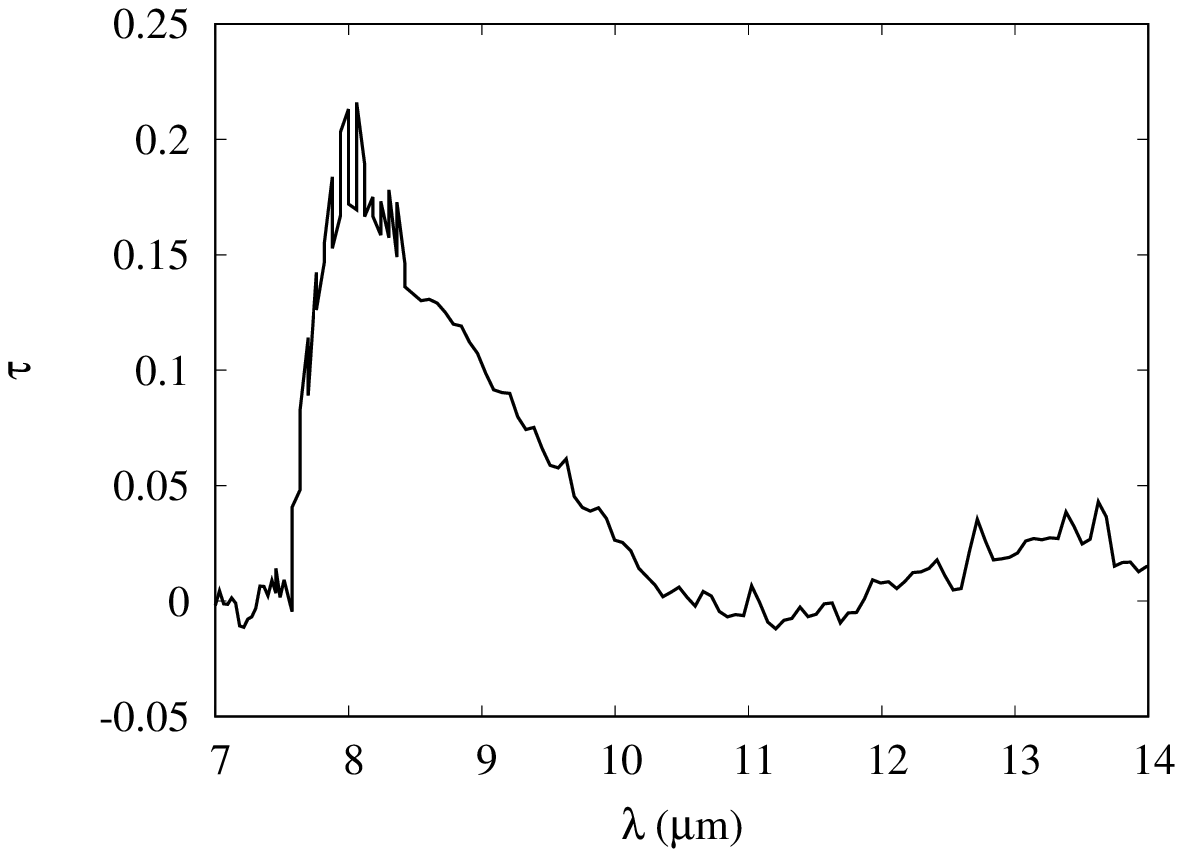}
\includegraphics[width=7.8cm,keepaspectratio]{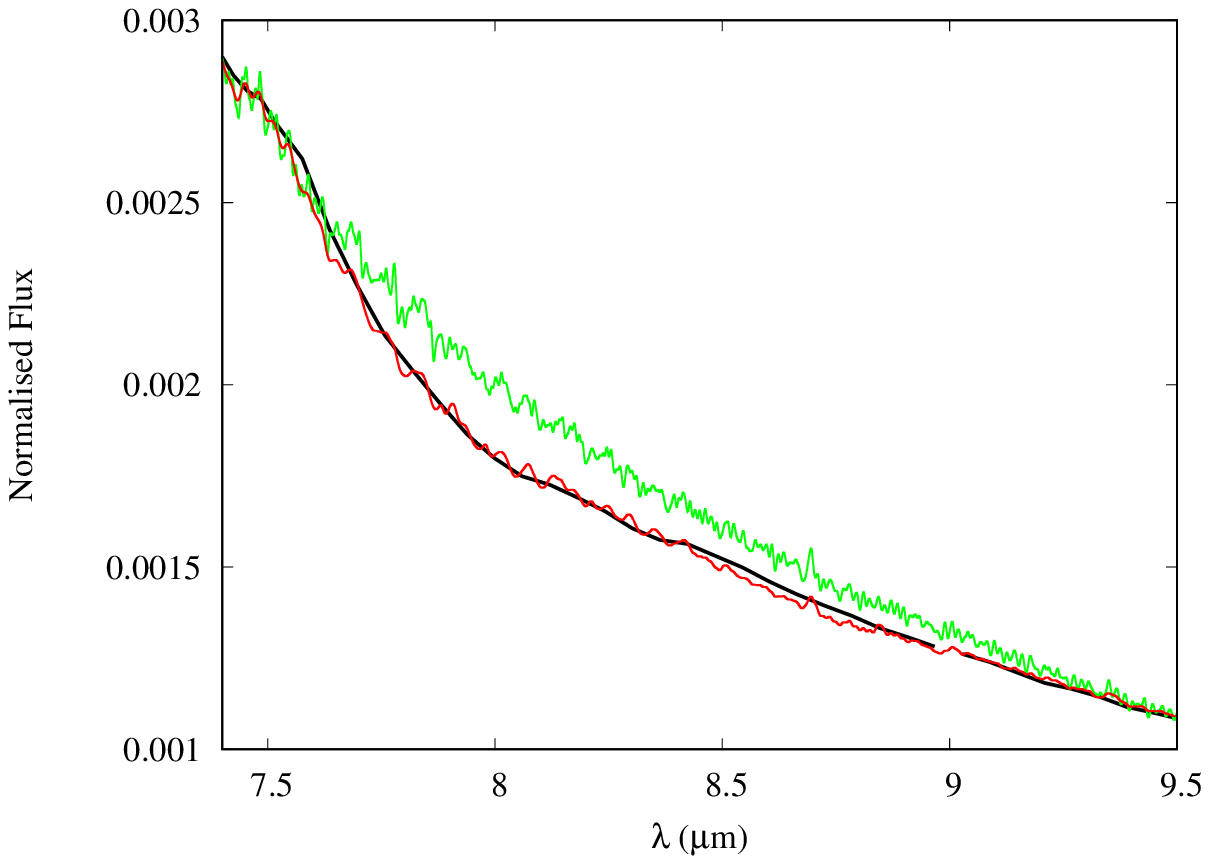}
\caption{Top: Power-law fit to the continuum around 8\mic\
as discussed in text.
Middle: {Optical depth in SiO feature.}
Bottom:~8\mic\ feature in \tcrb\ (black). Green curve is contribution of
photospheric SiO. Red curve is the combined effect of absorption by SiO
in the RG photosphere, with an additional contribution from SiO in the RG wind. 
{The SiO column density is 
$2.9\times10^{17}$~cm$^{-2}$, the wind temperature is 1,000~K.}
See text for details.
\label{SiO}}
\end{figure}

Clearly visible in the \sirtf\ spectrum is a weak absorption 
feature at around 8\mic\
(see Figs~\ref{tcrb_sp} and \ref{SiO}). To highlight this feature we 
have fitted a power-law function
 \[ f_\nu = A \, \lambda^\alpha   \]
to the continuum on either side of the 
feature (see Fig~\ref{SiO}); the fit shown is for $\alpha=-1.71\pm0.02$.
The wavelength of maximum absorption and the profile of this feature
is consistent with the $\Delta\upsilon=1$ fundamental transition in
gas-phase SiO, which has long been known in the spectra of
late giants and supergiants \citep*[see e.g.][]{geballe,decin,sloan}. 
We therefore attribute the feature in \tcrb\ to 
absorption by gas-phase SiO and, in view of the nature of the system, the 
SiO must originate in the M3III component.

SiO is added to the synthetic spectrum used in Section~\ref{1213} to 
model the IRTF
spectrum, except that the irradiation by the WD is ``switched off'' 
in view of the relative 
orientations of the WD, RG and observer (see Section~\ref{phase}). 
We take data for SiO (energy levels, partition function, etc.) from 
the {\it Exomol} project
\citep*{barton,hill}, which are appropriate for a zero pressure 
gas (i.e. pressure broadening is not taken into account); only the 
isotopologue $^{28}$Si$^{16}$O is included. 

We find that, for solar 
abundances, the optical depth in the photospheric SiO bands does 
not fully match the \sirtf\ 
observations: {\em additional absorption is required to accound 
for the depth of the feature}. 
We might make ad hoc adjustments to the abundances of Si and/or O 
in the RG atmosphere to enhance the SiO abundance, and hence 
strengthen the SiO fundamental 
feature. However such action would have adverse effects 
on other (atomic) 
features in the spectrum where the agreement between synthetic and 
observed spectra is good.

We conclude that the additional absorption must arise in the RG wind.

\subsection{SiO in the RG wind}
\label{wind}
We therefore amend the calculated flux to 
\[ F_\lambda = F_{\rm SAM12}  \: \exp(-\tau_{\rm SiO}) \:,\]
where $\tau_{\rm SiO}$ is the optical depth due to SiO in the RG wind. For 
simplicity the absorbing material is placed
in a plane-parallel slab between the RG and the observer, and the only free 
parameters are the column density of SiO and the temperature of the gas 
in the wind. 
We find that the entire SiO feature is well fitted by the 
photospheric component, but with an 
additional component having  $T\simeq1,000$~K and a column density of SiO 
{$N(\rm {SiO})=2.9\times10^{17}$~cm$^{-2}$;
this is comparable to the SiO column density in field RGs \citep{ohnaka}.}

If the absorbing column reported by \citeauthor{kennea} 
is mostly in the form of H, we estimate 
{that $N(\mbox{SiO})/N(\mbox{H})\sim1.3\times10^{-6}$;
the value in M-type Asymptotic Giant Branch (AGB) stars is in
the range $\sim[2-50]\times10^{-6}$ \citep{GD}, so its value in the wind
of \tcrb\ is very much at the lower end of this range.}

\begin{figure}
\includegraphics[width=7.8cm,keepaspectratio]{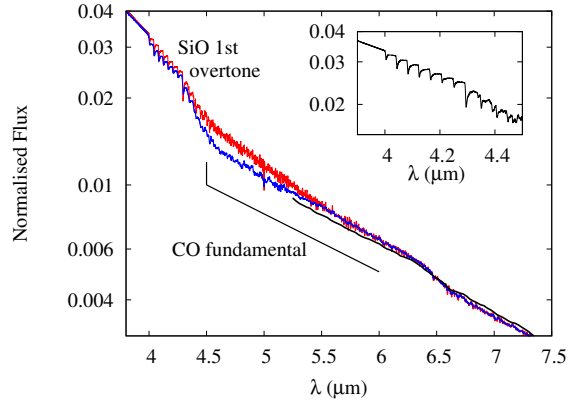}
\caption{Spectrum in the region of the CO fundamental and SiO 
first overtone. Black line: \sirtf\ IRS data. 
Red line: RG photosphere only.
{Blue line: RG photopshere plus column density
of $2.9\times10^{17}$~cm$^{-2}$ (SiO) and
$9\times10^{18}$~cm$^{-2}$ (CO).
Approximate extent of the $^{12}$CO fundamental is indicated.
Inset shows detail around SiO first overtone.}
\label{pred}}
\end{figure}

The binding energy of SiO (8.26~eV) is substantially lower than that of
CO (11.09~eV), so CO will be better able to survive in the 1,000~K RG 
wind than SiO. {Indeed the low value of $N(\mbox{SiO})/N(\mbox{H})$
deduced above may be related to the hostile environment in which the SiO 
is located in \tcrb.} As the column density of CO in the RG photosphere is 
$\simeq30\times$ 
greater than that of SiO, we estimate that the column 
density of CO in the wind should be 
{$\sim{9}\times10^{18}$~cm$^{-2}$.}
A spectrum in the region of the CO fundamental at 4.67\mic\ would 
therefore be particularly valuable because we predict that, as is the 
case with the SiO fundamental, the CO fundamental will be deeper than 
that corresponding to the first overtone\footnote{We note that, at 
$\sim1,000$~K, the relative populations of the $\upsilon=1$ and $\upsilon=2$ 
vibrational levels in CO is $\sim27:1$, so there is negligible contribution 
to the observed first overtone CO absorption from the wind.}
discussed in Section~\ref{1213}. 

There is some evidence that the CO fundamental is indeed deeper than 
that expected from a pure photospheric contribution (see Fig.~\ref{pred}, 
in which the SAM12 model is plotted to fit the region of the CO 
fundamental absorption). The CO fundamental is expected to cover
the wavelength range $\sim4.2-6$\mic, but unfortunately our Spitzer data 
only start at 5.2\mic. It can still be seen however that a
contribution from CO, over and above photospheric, is needed to
explain the depth of the CO fundamental in the Spitzer data. 
As discussed above, we believe that this additional CO 
originates in the RG wind and, based on the SAM12 fit in 
Fig.~\ref{pred}, we estimate that the column density of 
CO in the wind is $\sim{9}\times10^{18}$~cm$^{-2}$.
A spectrum covering the entire wavelength range covered by
the CO fundamental would be valuable to tie down this number.
A search for the corresponding SiO first overtone absorption 
(band-heads in the wavelength range 4.02--4.13\mic) would also 
be helpful, in particular high resolution spectroscopy to determine 
isotopic ratios; the predicted profiles of the SiO first overtone are 
shown in  Fig.~\ref{pred}.

We note here that a search for SiO maser emission from \tcrb\ in the  
$J=1\rightarrow0$, $\upsilon=1$ transition at 43.122~GHz yielded a 
non-detection \citep{blair}. However \cite{deguchi} detected SiO maser
emission, from both the $J=1\rightarrow0$, $\upsilon=1$ (43.122~GHz) and
$J=1\rightarrow0$, $\upsilon=2$ (42.821~GHz) transitions in V407~Cyg 
several weeks after its 2010 eruption. They found dramatic and complex
changes in the SiO maser emission from the Mira component as the 
post-eruption shock swept through the Mira wind.
In view of our detection of SiO in the RG wind of \tcrb,
an attempt to detect maser emission is highly desirable,
both before and after the anticipated RN eruption. Such
observations would provide a rare opportunity 
to study the dynamics of the nova eruption. 

\subsection{No dust in the \tcrb\ system}
\label{dust}
We see from Fig.~\ref{tcrb_sp} that, if there is a ``conventional'' 
9.7\mic\ silicate feature in the \sirtf\ IRS spectrum of \tcrb\ in 
either emission or absorption, its peak relative to the adjacent 
continuum is $<20$~mJy ($3\sigma$); we can place a similar limit 
on the absorption in/emission by various polymorphs of silica (SiO$_2$), 
which peak in 
the wavelength range $\sim8.4-10$\mic\ \citep{koike}. In stark contrast 
to the case of \rs\ \citep[see][]{evans1,woodward,rushton}, 
there is no evidence for silicate dust in the \tcrb\ system.

Why is gas-phase SiO detectable in the wind of \tcrb\ whereas 
the 9.7\mic\ solid state silicate feature is essentially absent,
while the silicate feature is in emission in \rs? On the face of 
it the two systems are similar, with
giant secondaries, similar orbital periods and massive WDs.

The conventional picture of dust formation in AGB stars 
\citep[see e.g.][]{HO} is that simple molecules (such as 
TiO, CO, SiO in the case of oxygen-rich systems) are present 
in the wind in a region just above the RG atmosphere.
Above this region is the dust-forming layer, where oxygen-rich species,
such as silicates, and aluminium and magnesium oxide solids,
form and grow.
However a necessary condition for the formation and growth of dust grains
is the presence of nucleation sites: if such sites are not present 
then dust can not form in the wind. 

In the case of \tcrb\ and \rs,
a key aspect in which they seem to differ is in their X-ray emission.
\rs\ is relatively X-ray faint, in contrast to \tcrb\ \citep[see][]{mukai}. 
Moreover, as noted in Section~\ref{tcrb}, \tcrb\ is a hard X-ray source 
whereas \rs\ is not. We suggest that the layer in the RG wind of 
\tcrb, where nucleation sites would usually be expected to form, is 
persistently exposed to hard radiation from the WD, 
thus preventing the formation of nucleation sites.

\subsection{The \fion{O}{iv} fine structure line}

Apart from the \pion{He}{i} $\lambda=1.0833$\mic\ triplet
(see Fig.~\ref{tcrb_sp}), there appear to be no particularly strong 
emission lines in the IR spectrum of \tcrb. 
Weak features seen are those of \pion{O}{i} $\lambda=0.8446\mic, 1.1287$\mic, 
and Pa-$\gamma$ and Pa-$\beta$ at 1.0940\mic\ and 
1.2818\mic\ respectively \citep[see][]{munari2}.

A notable absentee is the \fion{O}{iv} fine structure line at 
$\lambda=25.89$\mic\ (Fig.~\ref{tcrb_sp}), which is ubiquitous in
mature classical novae \citep[e.g.][and references therein]{basi,helton}, 
and is present in the IR spectrum of the RN \rs\ \citep{evans1,rushton}. 
We set a $3\sigma$ upper limit of 6.2~mJy on the peak flux density
in the \fion{O}{iv} line.

We suppose that this line is quenched by collisional de-excitation
by electrons. At $\sim1000$~K -- the temperature deduced for the SiO-bearing
material (see Section~\ref{wind}) -- the critical electron density
below which radiative de-excitation dominates collisional de-excitation
for the \fion{O}{iv} line is $4.64\times10^3$~cm$^{-3}$; 
at $10^4$~K, the critical density is $9.94\times10^3$~cm$^{-3}$ 
\citep[see Table~4 of][]{helton}.
The absence of the \fion{O}{iv} line in \tcrb\ implies therefore
that the electron density in the wind must be at least 
$\sim5\times10^3$~cm$^{-3}$.

It is not trivial to convert a lower limit on electron density
to a hydrogen column density; however this lower limit on the 
electron density is not inconsistent with the high column 
density required by the X-ray observations.

\section{Concluding remarks}

We have presented IR photometry and spectroscopy of the recurrent 
nova \tcrb. We summarise our conclusions as follows:
\begin{enumerate}
\item we have found the SiO fundamental -- commonly seen in evolved stars -- 
in absorption at 8\mic. The observed strength of the feature is such that a 
significant proportion must arise in the RG wind. The deduced SiO column 
density in the wind 
{is similar to that seen in the winds of field RG stars;}
\item a search for SiO maser emission from \tcrb, both before and after
its next eruption, is desirable;
\item we predict that the CO fundamental will also be stronger than is necessary 
to account for photospheric absorption, and that the column density of CO in
the wind is {$\sim{9}\times10^{18}$~cm$^{-2}$;}
\item the $^{12}$C/$^{13}$C ratio in the {RG is $\simeq5$,} very much at the 
low end of that expected on the basis of post-dredge up evolution;
\item there is no evidence of silicate dust in the environment of \tcrb's RG, 
in contrast to the case of \rs. This is likely a consequence of the stronger,
and harder, X-ray source in the \tcrb\ binary compared with that in \rs.
\end{enumerate}

When the next eruption of \tcrb\ does occur, the ejected material will run 
into and shock the RG wind, so we might expect that the evolution of 
the eruption will, superficially at least, resemble those of the eruptions 
of \rs, with strong X-ray emission and 
strong coronal line emission;
indeed  coronal lines were prominent in the optical spectrum of \tcrb\
during the 1946 eruption \citep*{bloch1,bloch}.
A list of IR coronal lines is given in Table~2 of \cite{basi}; many of these
lines will be accessible to the FORCAST instrument on \sof.

Moreover, the recent detection of classical novae as $\gamma$-ray sources
\citep{ackermann}, and the nature of the \tcrb\ system, strongly suggests
that \tcrb\ will likely be a $\gamma$-ray source when it erupts
\citep{banerjee,metzger1,metzger2}. 

Clearly, suitable observations of \tcrb\ when it does eventually erupt are 
highly desirable, and we hope that the present work will provide an impetus to 
further observations of \tcrb\ in quiescence as we approach the next eruption.

\vspace{-4mm}

\section*{Acknowledgments}

{We thank the anonymous referee, Dr Tom Geballe and the editor 
for their help with improving the original version of this paper.}

Based in part on observations made with the NASA/DLR Stratospheric Observatory for Infrared Astronomy (\sof). \sof\ is jointly operated by the Universities Space Research Association, Inc. (USRA), under NASA contract NNA17BF53C, and the Deutsches SOFIA Institut (DSI) under DLR contract 50~OK~0901 to the University of Stuttgart.

This publication makes use of data products from the Two Micron All Sky Survey, which 
is a joint project of the University of Massachusetts and the Infrared Processing and 
Analysis Center/California Institute of Technology, funded by the National Aeronautics 
and Space Administration and the National Science Foundation

The Infrared Telescope Facility (IRTF) is operated by the University of 
Hawaii under contract NNH14CK55B with the National Aeronautics and Space Administration.
We are very grateful to David Sand for making possible the IRTF observation.

The research work at Physical Research Laboratory is supported by the 
Department of Space, Government of India.
RDG was supported by NASA and the United States Air Force. 
SS is grateful for partial support from NASA and HST grants to ASU.
CEW was supported by NASA \sof\ resources under contracts from USRA.
UM is partially supported by the PRIN-INAF 2017 ``Towards the SKA
and CTA era: discovery, localisation and physics of transient sources'' (PI M. Giroletti)

\label{lastpage}
\end{document}